\documentclass[12pt,preprint]{cmlpasp}
\usepackage{multirow}
\newcommand{\beq}{\begin{equation}}
\newcommand{\eeq}{\end{equation}}
\newcommand{\beqn}{\begin{eqnarray}}
\newcommand{\eeqn}{\end{eqnarray}}

\begin{document}
\title{Searching for Possible Exoplanet Transits  
from BRITE Data through a Machine Learning Technique 
}
\author{Li-Chin Yeh$^{1}$ and Ing-Guey Jiang$^{2}$}

\affil{
{$^{1}$Institute of Computational and Modeling Science,}\\
{National Tsing Hua University, Hsin-Chu, Taiwan}\\
{$^{2}$Department of Physics and Institute of Astronomy,}\\
{National Tsing Hua University, Hsin-Chu, Taiwan} 
}
\email{lichinyeh@mx.nthu.edu.tw, jiang@phys.nthu.edu.tw}

\begin{abstract} 

The photometric light curves of BRITE satellites were examined through 
a machine learning technique to investigate whether there are possible 
exoplanets
moving around nearby bright stars. Focusing on different transit periods, 
several convolutional neural networks were constructed to search for 
transit candidates. The convolutional neural networks were trained with 
synthetic transit signals combined with BRITE light curves until the 
accuracy rate was higher than 99.7 $\%$. Our method could efficiently 
lead to a small number of possible transit candidates. 
Among these ten candidates, two of them, HD37465, and HD186882 systems,
were followed up through future observations with a higher priority. 
The codes of convolutional neural networks employed in this study 
are publicly available at 
http://www.phys.nthu.edu.tw/$\sim$jiang/BRITE2020YehJiangCNN.tar.gz.
\end{abstract}

\section{Introduction}

The discoveries of extra-solar planets (exoplanets) 
have significantly impacted  astronomical sciences 
and advanced our understanding of the universe.
Among many detection techniques, the transit method
became one of the major players after the Kepler Space Telescope
detected thousands of exoplanet candidates (Borucki et al. 2011). 
The light curves obtained through the Kepler Space Telescope
(Kepler light curves hereafter)    
have been released.  These valuable data were further examined
and led to many potentially useful results. For example, Xie (2013) 
did the transit timing analysis on Kepler light curves
and found several near-resonance planetary pairs.
Ioannidis et al. (2014) further analyzed  the 
light curves of Kelper-210 and detected two short-period planets.
Later, Sun et al. (2019) employed the light-curve data
of the Kepler-411 system and revealed the existence of additional planets.
Thus, it is now suggested there are four planets in the Kepler-411 system.
Recently, Socia et al. (2020) discovered from Kepler light curves,
a new Neptune-sized transiting circumbinary planet.

To monitor more stars, the telescopes of survey projects need 
to be capable of measuring the photometry of faint stars
with larger apparent magnitude.   
The Kepler Space Telescope can monitor stars with an apparent magnitude around
$V=11 \sim 14$. Additionally, the CoRoT project focuses on stars with an 
apparent magnitude around $V=12 \sim 15$. Finally,
HATNet and WASP projects monitor stars with apparent magnitude 
$V=9 \sim 12$. 
These projects have detected many transit exoplanets,
and the fraction of stars hosting a planet is consistent 
with the reuslts suggested by Jiang et al. (2010). 
However, the brighter stars with $V < $ 5 mag
remain unexplored. It is not clear
yet whether or not there are any exoplanets orbiting around these 
brighter stars. 
{\it In particular, when the brighter stars are nearby, 
it is very interesting to investigate 
whether there are any habitable exoplanets moving around them.}

BRITE (BRIte Target Explorer) is a space-based project 
with five operating nano-satellites that 
are equipped with telescopes of 3 cm aperture (Weiss et al. 2014). 
The main task of BRITE is to monitor 
bright stars, and the photometry precision is around mmag for stars brighter 
than $V \sim$ 5 mag. It has 
studied oscillation modes of many pulsating stars. For example, 
BRITE photometry detected several new modes 
for the $\beta$ Cep star HD29248 (Handler et al. 2017). 
This star now has ten low-order modes and 
seven high-order gravity modes. Potentially, BRITE could also pose 
some observational constraints 
on the number of exoplanets hosted by nearby bright stars. 
Passegger and Weiss (2006) estimated the number 
of exoplanet transits which could be detected by BRITE for 
the main sequence stars brighter than $V = 4$ mag. 
They concluded that a few Neptune-sized and Jupiter-sized exoplanets 
shall be detectable. Mol Lous et al. (2018) was the first to search 
for transiting planets in $\beta$ Pictoris system from the BRITE data because 
this system was well monitored by BRITE and might detect additional planets 
given that a Jupiter-mass planet,
$\beta$ Pictoris b, was directly imaged in 2009 (Lagrange et al. 2009, 2010). 
Although Mol Lous et al. (2018) did not report a positive result, 
this gave us the impetus to examine the BRITE data 
through a new transit searching method such as the machine-learning technique.

The conventional method used for the transit detection task is to 
calculate some important quantities from the data, and then criteria 
are set to automatically classify them. For complicated cases, one might 
need to make several levels of judgment and build up a pipeline 
to deal with the data. 
The main advantage of machine learning is that the computer code gets 
modified and can be improved by 
itself through a learning process. The conditions of classification 
can be changed, thus, improving the 
capability of making correct judgments.

Machine learning has been employed to investigate many astronomical problems 
in recent years (Longo et al. 2019). It was used to classify the morphology 
of galaxies (Perez-Carrasco et al. 2019) and 
the spectrum of stellar objects (Passegger et al. 2020) and to model 
stellar oscillation modes.
In particular, Hendriks and Aerts (2019) employed a deep learning technique 
to carry out asteroseismic modeling of the $\beta$ Cep star HD29248 
and found it to be a rather evolved star.

In fact, machine learning is already in use as a new tool to vet 
or detect exoplanets. Pearson et al. (2018) 
showed a triage to employ machine learning to search for new exoplanet transits. They compared the results 
obtained through several methods. Likewise, Shallue $\&$ Vanderburg (2018) demonstrated that the machine 
learning technique, particularly the convolutional neural network (CNN), can be useful to vet and confirm 
threshold crossing events and lead to the discoveries of new exoplanets. Recently, Ansdell et al. (2018) 
suggested to include stellar parameters, etc. into CNN structures for improved accuracy. 
Later, Dattilo et al. (2019) employed CNNs to identify two previously unknown exoplanets from the K2 
mission of the Kepler Space Telescope. Yu et al. (2019) attempted to use CNNs to vet possible transits 
from the newly released data of the Transiting Exoplanet Survey Satellite (TESS).

Considering the above-mentioned developments, in this study, we planned to employ the CNN of deep learning 
to examine whether there are exoplanet transit candidates in BRITE data. As BRITE mainly 
monitors bright stars, we expected to gain some statistical data about exoplanets around these 
bright stars, with an aim to improve our understanding of the general picture of the overall exoplanet 
population.

Machine learning and CNN are briefly summarized in \S 2. 
The observational BRITE data and the 
preparation of CNN input light curves are presented in \S 3. 
The training of our CNN models is described 
in \S 4, the results are presented in \S 5, 
followed by conclusions in \S 6.

\section{Machine Learning with Convolutional Neural Network} 

In  machine learning, when some data is given as an input to a computer code, the code does some calculations 
and produces results that are taken as a feedback to improve the computer code itself. Deep learning is a 
type of machine learning technique that imitates the human brain and consists of many calculation units 
called neurons. Several neurons are assigned to one layer. Usually, a number of layers are needed and 
they are connected to form a neural network. When the input data of a layer is designed to do the 
convolution with a filter, i.e. kernel, this layer is called a convolutional layer. When a neural 
network consists of convolutional layers, it is called a CNN. An introduction of CNN can be found 
in Chintarungruangchai $\&$ Jiang (2019).

In this study, we have considered the one-dimension convolutional neural network (1D-CNN) of machine learning 
techniques. With this deep-learning technique, we decided to examine the possible transits with periods 
from one to five days. One way to proceed was to train a CNN model through synthetic light curves with the 
above transit periods and use this CNN model to detect possible transits from data. However, to make our 
CNN models more sensitive to the transit signals with specified periods, we decided to have four different 
CNN models that focus on different periods. Thus, there were \\
(a) Model P12: the detection of possible transits with periods between one and two days; \\
(b) Model P23: the detection of possible transits with periods between two and three days; \\
(c) Model P34: the detection of possible transits with periods between three and four days; \\
(d) Model P45: the detection of possible transits with periods between four and five days. \\
The architecture of the above 1D-CNN models contains four convolutional layers. 
For each convolutional layer, the channel number, kernel size, 
and stride were set as 64, 5, 1, respectively.
In addition, when we trained Model P12, only those samples were employed which injected transit signals 
with periods between one and two days. Similarly, Model P23, Model P34, Model P45 were also trained by 
the light curves with corresponding transit periods.

\section{The Light Curves}

In this section, the BRITE observational light curves are described. In addition, the procedure to inject 
transit signals into light curves, and the details of light-curve folding are also presented.

\subsection{BRITE Data}

   For this study, all BRITE data sets of Data Release 2 (R2) and Data Release 3 (R3) were retrieved from 
BRITE Public Data Archive. These data sets were obtained by five BRITE nano-satellites: blue-filter 
BRITE-Austria (BAb), red-filter BRITE-Heweliusz (BHr), blue- filter BRITE-Lem (BLb), red-filter 
BRITE-Toronto, and red-filter Uni-BRITE (UBr). The name of the monitored star, the observational duration, 
and the names of the employed nanosatellite name are clearly mentioned in the header of each data file. 
For one particular star, there could be more than one data file. In this case, we would check which 
nano-satellite was used to obtain the measurements in these data files. 
If the observations were done by 
different nano-satellites, they were considered to be separate light curves. If the photometric measurements
in several files were done through the same nano-satellite, these measurements were combined to become one 
light curve.

   Our goal was to detect possible transit candidates from a large number of light curves, therefore, to be 
efficient, we planed to use a simple procedure to reduce the scattering of light curves without going 
through elaborate corrections (Popowicz et al. 2017). Thus, we chose suitable intervals to bin the data to 
average out the scattering. We decided to calculate the average flux within a 20-minute bin. This choice of 
20 minutes can maximize the scattering reduction but still would have several data points for one transit 
event because a typical transit duration is about 60 to 200 minutes. The above-binned data was normalized 
with respect to the averages of observational runs. Then, the standard deviation was determined and the 
$3\sigma$ outlier data was removed. All light curves were obtained in this study through the above 
procedure before any further analysis. They were referred to as BRITE {\it light curve} hereafter. 
Fig. 1 shows one such BRITE light curve of the star HD37468 as an example. 
All the BRITE data sets used in this study are listed in Table 1. In total, there are 35 BRITE light curves.

\begin{table}
\centering
\begin{tabular}{|c|c|c|c|c|} \hline
 Satellite &  Data ID & Star & Number of Data Points (N)  &  Standard Deviation ($\sigma$) \\ 
\hline
 BAb &(BAb-1)  & HD37043    &   1385  & 0.00444  \\ \cline{2-5}
  &(BAb-2) &HD121263     &  1870 & 0.00388  \\ \cline{2-5}
 &(BAb-3) &HD35468    &1380  &0.00279 \\ \cline{2-5}
  &(BAb-4)& HD33111     &1116 &0.00435  \\ \hline
  BHr & (BHr-1)& HD37468  &  2022& 0.00452 \\ \cline{2-5}
 & (BHr-2) & HD35468    &  1995 &  0.0049  \\ \hline
 BLb &(BLb-1)& HD35468  &  1842 &  0.00377   \\ \cline{2-5}
 &(BLb-2)& HD186882    &  1574&  0.0039   \\ \hline
BTr & (BTr-1) &HD80404   &   1397&   0.00342\\ \cline{2-5}
 & (BTr-2)& HD34503   &   1382&  0.00422\\ \cline{2-5}
 & (BTr-3)&HD79351     &  1387&  0.00467 \\ \cline{2-5}
 & (BTr-4)&HD198183  &   1569&  0.00449  \\ \cline{2-5}
 & (BTr-5)&HD74772   &  1399&  0.00284\\ \cline{2-5}
 & (BTr-6)&HD82434     &  1406&  0.00371\\ \cline{2-5}
 & (BTr-7)&HD81188    &  1388 & 0.00411  \\ \cline{2-5}
 & (BTr-8)&HD74006   &  1396& 0.00304\\ \cline{2-5}
 & (BTr-9)&HD64740      &  1400& 0.00463   \\ \cline{2-5}
 & (BTr-10)&HD33111     &  1380&  0.00336\\ \cline{2-5}
 & (BTr-11)&HD68553   &  1391&  0.00496\\ \cline{2-5}
 & (BTr-12)&HD30652    &  1370&  0.00407\\ \cline{2-5}
 & (BTr-13)&HD202444  &   1224&  0.00225\\ \cline{2-5}
 & (BTr-14)&HD35468    & 1378& 0.00370\\ \cline{2-5}
 & (BTr-15)&HD76728      &  1375& 0.00468 \\ \cline{2-5}
 & (BTr-16)&HD64440     &  1399&  0.00252\\ \cline{2-5}
 & (BTr-17)&HD78004   &  1374&  0.00455\\ \cline{2-5}
 & (BTr-18)&HD71129   &   1395&   0.00384\\ \cline{2-5}
 & (BTr-19)&HD36512   &  1380& 0.00485 \\ \cline{2-5}
 & (BTr-20)&HD74575    &  1082&  0.00392\\ \cline{2-5}
 & (BTr-21)&HD186882   &   1222&  0.00245\\ \hline
UBr & (UBr-1)& HD20902  &  3742&  0.00440 \\ \cline{2-5}
 & (UBr-2)& HD23630     &  3736& 0.00478\\ \cline{2-5}
 & (UBr-3)&HD121263  &  2648& 0.00426 \\ \cline{2-5}
 & (UBr-4)&HD33111      & 1413& 0.00395\\ \cline{2-5}
 & (UBr-5)&HD134505   &  2646&  0.00491\\ \cline{2-5}
 & (UBr-6)&HD35468    &  1405& 0.00468\\ \hline
\end{tabular}
\caption{BRITE light curves}
\end{table}

\subsection{Injecting Transit Signals}

  To search for transit candidates from light curves, the CNN model needs to be trained by both 
non-transit and transit light curves. The BRITE light curves were used as the training samples of 
non-transit light curves. The transit signals were injected into the BRITE light curves to produce 
synthetic transit light curves to be used as training samples.

  However, there may be real transits in BRITE light curves, and thus they cannot be used as non-transit 
training samples. As presented in the next subsection, the BRITE light curves were folded with a huge 
number of assumed trial periods. For almost all cases, the possible real transit signals were destroyed 
during the folding process. The probability that the real transit is not destroyed in folded light curves 
is extremely small. A huge number of training samples were employed, this tiny defect did not affect 
the training results of CNN models. Thus, the folded BRITE light curves were treated as non-transit 
light curves in training samples.

  The transit signals were modeled by the method in Mandel and Agol (2002). The parameters included the 
transit period ($p$), the ratio of the orbital semi-major axis to the stellar radius ($a_{rm os}$), 
the ratio of planet radius to the stellar radius ($R_{rm ps}$), the orbital inclination ($i$), 
the linear limb darkening coefficient ($u_1$), and the quadratic limb darkening coefficient ($u_2$). 
As described in the previous section with the four CNN models which correspond to different ranges of 
transit periods, the parameter $p$ was set to be uniformly distributed between one and two days for Model P12, 
two and three days for Model P23, three and four days for Model P34, four and five days for Model P45. 
Other parameters are set to be uniformly distributed within the intervals shown in Table 2.

\begin{table}
\centering
\begin{tabular}{c l}  \hline
parameter &  range  \\ \hline
$a_{\rm os}$  & 10-30 \\
$R_{\rm ps}$ & 0.06-0.15 \\ 
$i$ & 86-90 (degrees)\\
$u_1$  & 0.5 \\
$u_2$ & 0.0 \\ \hline
\end{tabular}
\caption{Parameters of transit signals}
\end{table}

   The transit light curves after injecting these transit signals into the BRITE data are presented as Fig. 2. 
For convenience, the transit period ($p$) was set at 1 day for the results in Fig. 2. Fig. 2(a)-(c) 
present the outcomes injecting the transit signal into the BRITE light curve of the star HD68553, and 
Fig. 2(d)-(f) present the outcomes for the star HD202444. From these plots, it is clear that 
the transit depth is much more sensitive to the parameter $R_{\rm ps}$.

\subsection{The Folding}

   BRITE is not designed for the detection of exoplanets, so the time for data resolution and monitoring 
duration is very different from usual transit observations. 
The main characteristics of BRITE data are 
the high time resolution within intervals of about 15 minutes 
and various gaps between these intervals. Thus, 
the folding is the best way to reveal the possible transit signals.

  However, folding can enhance the signals when the folding period is the same as the signal period and 
can destroy signals when the folding period is different from the signal period. Therefore, 
we had to conduct folding for different trial periods, and a huge amount of folded light curves 
could be produced. The CNNs of machine learning then play a very important role in judging 
which ones are transit candidates.

To enable the CNN models to detect transits from 
those light curves folded with slightly different periods, 
we also carried out folding with periods that were two-minute 
shorter and two-minute longer than the 
transit period, i.e. $p \pm 2/1440$ (day). 
In Fig. 3, we present both transit light curves and 
non-transit light curves of two stars after folding. 
The folding periods for the star HD68553 
in Fig. 3(a)-(c) were set to be 1438 minutes, 
1440 minutes, and 1442 minutes, respectively. 
Similarly, the folding periods for the star HD202444 
in Fig. 3(d)-(f) were set to be 1438 minutes, 
1440 minutes, and 1442 minutes, respectively. Other transit parameters include 
$a_{\rm os}=20.0$, $R_{\rm ps}=0.07$, $i=88$, and period 
$p=1 {\rm (day)}=1440{\rm (min)}$  for all light curves in Fig. 3.

\section{The Training Processes}

     For a given number of light curves $N_t$ (which is usually larger than 10000), 80$\%$ of these are 
used as the training data, 10$\%$ include the validation data, and the remaining 10$\%$ are the 
testing data. Note that half of all the above light curves are transit light curves and half 
are non-transit light curves. The training data was separated into subsets, called mini-batches. 
We randomly assigned 16 light curves to each mini-batch (except the final mini-batch which had fewer 
light curves). Each mini-batch was fed into the CNN model one by one until all were used. When all 
training data are used, it is called one epoch. Before the next epoch starts, the light curves in 
mini-batches are randomly assigned again. We observed many epochs, and every time right after one 
epoch was completed, the CNN model was used to predict the answers of the validation data. 
The percentage of the correct answers was defined to be ``accuracy'', and each epoch had a value of accuracy. 
The stopping condition of the training process was indicated when the difference of two successive 
accuracies was less than 0.001 and the number of epochs was larger than 25. 
Once the training process stopped, the CNN model settled and could be used to predict the answers 
of the testing data and obtain the corresponding accuracy.

\subsection{The CNN Models}

   Here we describe how we obtained the CNN models, i.e. Model P12, P23, P34, and P45, which were used to
search for possible transits. Taking Model P12 as an example, we first needed to inject transit signals 
into BRITE light curves. To decide a transit signal to be injected, we chose a period $p$ through a uniformly 
distributed random number ranging between one and two days. Concurrently, other parameters $a_{\rm os}$, $R_{\rm ps}$, and $i$  were also randomly chosen from uniform distributions within the intervals listed in Table 2.

   Each light curve was folded with the same chosen period $p$. When $p$ is between one day and two days, 
the number of data points of one light curve is between 72 and 144 (because the bin size was set as 20 min). 
To make the Model P12 always have the same number of input nodes, we used interpolation to 
generate 144 data points for all light curves and set the number of 
input nodes as 144 for Model P12. 
Each light curve was also folded with a period that is two-minute larger and two-minute smaller than $p$. 
Thus, three folded light curves were generated from one BRITE light curve. These three were injected 
with the chosen transit signal and generated another three folded light curves with transit. Therefore, 
a total of six folded light curves were generated from one BRITE light curve to become a standard 
input with 144 nodes for Model P12.

  The typical number of training samples is usually larger than 10000 for the construction of any CNN models 
(Pearson et al 2018). Thus, we used a random number to decide which BRITE light curves could be used 
to generate six folded light curves. All our 35 BRITE light curves were candidates to be picked and 
this is repeated until there are several $N_s$ input light curves ($N_s$=6000).

  Each of these generated input light curves was used as training samples following the process 
previously described. The whole process was repeated five times to get five Model P12. 
Other CNN models, such as Model P23, Model P34, and Model P45 were also produced in the same way. 
In Table 3, the range of periods, input node numbers, and the accuracies are listed. 
Besides the accuracy, the results of reliability (the percentage of real transit light curves 
among those light curves which predicted to have transits) and completeness (the percentage of 
the light curves that are successfully predicted to have transits among those transit light curves) 
are also presented in Table 3. Because all CNN models were trained five times, five versions of 
Model P12, Model P23, Model P34, Model P45 could be generated. 
The values of accuracy, reliability, and completeness represent the corresponding median values, and 
the errors are their standard deviations of five versions. 
All five versions of these CNN models were employed to search for possible transits.

\begin{table}
\centering
\begin{tabular}{c l c c c c}  \hline
Model  & period (day) & Input Node Number & accuracy ($\%$)  & reliability ($\%$)  & completeness ($\%$)   \\ \hline
P12&$p=1 \sim 2$ &144 & 99.76$\pm$ 0.069  &99.81$\pm$ 0.195 & 99.68$\pm$ 0.077 \\
P23&$p=2\sim 3$ &216 &  99.90$\pm$ 0.030  &99.94$\pm$ 0.079 & 99.95$\pm$ 0.042 \\ 
P34&$p=3\sim 4$ &288 & 99.85 $\pm$ 0.026 & 99.78$\pm$ 0.093 & 99.90$\pm$0.048\\
P45&$p=4\sim 5$ &360  &  99.78$\pm$ 0.033 &99.99$\pm$ 0.056 & 99.63$\pm$0.093 \\  \hline
\end{tabular}
\caption{CNN models}
\end{table}

\subsection{The Learning Curve}

The accuracy of a CNN model is proportional 
to the sample size but it may get saturated 
when the sample size is big enough. 
Thus, to determine the best sample size,
it is important to produce the learning curve, i.e. the accuracy 
of a CNN model as a function of sample size.
With $N_s=6000$, six different total numbers of light curves,
$N_t$ = 2$N_s$, 4$N_s$, 8$N_s$, 16$N_s$, 32$N_s$, and  64$N_s$, were used to
train Model P12 to obtain the learning curve here. 

For each sample size $N_t$,  five cycles of training, validation, and testing were done
with different sample sizes. Finally, five accuracies were obtained.
The median value is defined as the accuracy of a particular $N_t$,
and the standard deviation gives the error bar. 
The accuracy as a function of $N_t$, i.e. the learning curve,
is presented in Fig. 4. From Fig. 4,  
we oberve that the accuracy is stable when the sample size 
is $N_t=32 N_s$, so the total number of employed light curves is 
$N_t=32 N_s$,  where $N_s=6000$, for the construction of CNN models. 

\section{The Results}

The results of the detection of transit candidates through CNN models 
are described here. We found a number of transit candidates and discuss 
two most likely transits that were modeled. 

\subsection{Transit Detection}

   After five versions of Model P12 were obtained, they were used to search for possible transits 
with periods between one and two days from our 35 BRITE light curves. To be ready as an input 
of Model P12, each BRITE light curve was folded with a folding period $fp = 1440 + j$ minute, 
where $j = 0, 2, 4, 6, 8, ..., 1440$. Only those detected by all five versions of 
Model P12 were considered as the transit candidates here. Then, Model P23, Model P34, 
and Model P45 were employed to search for possible transits with larger periods in the same way.

   The folding periods, in minutes of all the transit candidates detected by our CNN models, 
are listed in Table 4. The first column of Table 4 represents the Data ID of BRITE light 
curves originally assigned in Table 1. The second column is the transit period 
detected by Model P12. The third, 4th, and 5th columns are the transit periods detected 
by Model P23, Model P34, and Model P45, respectively. For example, for light curve BAb-1, 
three transits were detected. They represent transits with periods of 2160 minutes, 2892 minutes, 
and 6338 minutes. These three transits are put in different rows as they correspond 
to different periods. For light curve BAb-4, only one transit was detected. 
However, many more transits were detected for the light curve BHr-1. In particular, 
the transit with a period of 1714 minutes was detected repeatedly by three CNN models, 
Model P12, Model P23, and Model P34. In addition, for the light curve BLb-2, the transit 
with a period of 1440 min was also detected by more than one CNN models. 
Therefore, among all detected transit candidates, these two have larger probabilities to be real transits. 
However, one must be cautioned that 1440 min is exactly one day. 
Though BRITE is a space instrument, it might still suffer from various issues related to stability. 
Therefore, the transit for 1440 min could be a false one. The confirmation is out of 
the scope of this paper and is left for future further studies.
The other transit candidates were only detected, possibly due to week signals, 
by one CNN model and we decided to leave them to be investigated 
when there are more high-precision data available in the future.


\begin{table}
\centering
\begin{tabular}{|c| c| c| c| c|}  \hline
Data ID & Model P12 & Model P23 & Model P34 & Model P45\\  \hline
(BAb-1)  &  2160 &\multicolumn{3}{c|}{}  \\ \cline{2-5}
&  & 2892  &\multicolumn{2}{c|}{}   \\ \cline{2-5}
 &\multicolumn{3}{c|}{}      & 6338 \\ \hline
(BAb-4)   &\multicolumn{2}{c|}{}   & 5704 & \\ \hline 
(BHr-1)  & 1714 & 1714$\times$2 & 1714$\times$3 & \\ \cline{2-5}
& 1716   &\multicolumn{3}{c|}{}  \\ \cline{2-5}  
&  &4220 &\multicolumn{2}{c|}{}  \\ \cline{2-5}
 &\multicolumn{2}{c|}{} &  5144  &  \\ \cline{2-5}
 &\multicolumn{2}{c|}{} &5146 & \\ \hline
(BLb-1) & 1456  &\multicolumn{3}{c|}{} \\ \hline
(BLb-2) &  1440& & 1440$\times$3& \\ \cline{2-5}  
& 1444         &\multicolumn{3}{c|}{}  \\ \cline{2-5}
& 2868         &\multicolumn{3}{c|}{}\\ \cline{2-5}
& 2870         &\multicolumn{3}{c|}{}\\ \cline{2-5}
& 2872         &\multicolumn{3}{c|}{}\\ \cline{2-5}
& 2876        &\multicolumn{3}{c|}{}\\ \cline{2-5}
& 2878       &\multicolumn{3}{c|}{} \\ \cline{2-5}
&& 4312  &\multicolumn{2}{c|}{} \\ \cline{2-5}
&&4314   &\multicolumn{2}{c|}{} \\ \cline{2-5}
&& 4316 &\multicolumn{2}{c|}{}  \\ \cline{2-5}
&&4318  &\multicolumn{2}{c|}{} \\ \cline{2-5}
 &\multicolumn{2}{c|}{}& 4324 &  \\ \cline{2-5}
 &\multicolumn{3}{c|}{} & 5772\\ \cline{2-5}
 &\multicolumn{3}{c|}{}& 7188\\ \cline{2-5}
 &\multicolumn{3}{c|}{}&7196\\ \hline
(BTr-2) &\multicolumn{2}{c|}{}    & 4324&  \\ \hline
(BTr-12)  &\multicolumn{2}{c|}{}      & 4624& \\ \cline{2-5}
 &\multicolumn{3}{c|}{} &6986\\ \cline{2-5}
 &\multicolumn{3}{c|}{}& 7174 \\ \hline
(BTr-15)  &\multicolumn{2}{c|}{}       & 5014&  \\ \hline
(BTr-19)&  & 4230   &\multicolumn{2}{c|}{}  \\ \cline{2-5}
 &\multicolumn{2}{c|}{}& 4324 & \\ \cline{2-5}
 &\multicolumn{2}{c|}{}&4722 & \\ \hline
(BTr-20)&2656   &\multicolumn{3}{c|}{}  \\ \hline
(UBr-5) & 2870 &\multicolumn{3}{c|}{}  \\ \cline{2-5}
&2872  &\multicolumn{3}{c|}{} \\ \hline
\end{tabular}
\caption{The detected transit candidates}
\end{table}

\subsection{HD37468 and HD186882}

    Let us discuss these two systems whose light curves exhibit real detected transits here. 
The star HD37468, which gives light curve BHr-1, is also named as $\sigma$ Ori. 
It is actually a system with several stars. Among these stars, $\sigma$ Ori A and $\sigma$ Ori B 
form a visual binary with an orbital period of about 160 years. In addition, $\sigma$ Ori A is 
a binary star itself, comprising $\sigma$ Ori Aa and $\sigma$ Ori Ab, and their orbital period 
is 143 days. The detected 1714 min photometrical variation is definitely not due to 
the orbital motion of the binary $\sigma$ Ori Aa and $\sigma$ Ori Ab.

    These three stars appear very close to each other (within 0.26 arcsec) in the sky, 
and are not resolved by the BRITE telescope. Considering the stability, there may be 
a certain level of mass difference between three stars, which must be the most massive 
one which dominates the observed flux. It is likely that the 1714 min periodical variation 
could be due to an exoplanet orbiting around the most massive star.

    Though the system seems to be complicated, the photometric transit parameters 
could still be obtained as an estimation, assuming that the most massive star dominates 
the observed flux and the planet is moving around the most massive one. Here the model 
in Mandel $\&$ Agol (2002) is employed to fit the observational light curve through 
a Markov-Chain Monte-Carlo (MCMC) sampling process. 
Note that the limb-darkening effect is ignored for the occultation part of the light-curve model. 
The orbital period was fixed at 1714 minutes, the inclination was fixed to $i=90$, 
the limb-darkening coefficients were fixed to the values mentioned in Table 2. 
Other five parameters, including the mid-transit time ($t_{\rm tra}$), 
the mid-occultation time ($t_{\rm occ}$), the ratio of the orbital semi-major axis to 
the stellar radius ($a_{\rm os}$), the ratio of planetary radius to the stellar radius ($R_{\rm ps}$), 
and the ratio of planetary flux to stellar flux ($f_{\rm ps}$) were treated as MCMC samples.

   The MCMC sampling code {\it  MC3}, provided by Cubillos et al. (2017),
was used to generate seven chains of $10^{5}$ samples.
The best-fit model is then obtained with 
a value of reduced chi-square $\chi^2_{\rm red}=2.39$.
In Fig. 5, the observational data is shown as points
and the best-fit model is presented as the solid line.
Fig. 6 presents the two-dimensional 
and one-dimensional projections of MCMC posterior distributions, where
the dashed lines indicate the best-fit model.

    The star HD186882, which gives light curve BLb-2,
is also named as $\delta$ Cyg. It is a binary system with 
components $\delta$ Cyg A and  $\delta$ Cyg B.  Their separation is about
157 AU and they orbit around each other with a period 780 years.
In the sky, their angular separation is about 3 arcsec, thus they cannot
be resolved by the BRITE telescope. 
Their combined apparent magnitude was observed to be nearly 2.87. 
The primary star $\delta$ Cyg A is much brighter than the companion star
$\delta$ Cyg B. The detected transit with a period of 1440 min
could be due to a close-in exoplanet or a sub-stellar object
orbiting around $\delta$ Cyg A. 

The photometric transit parameters could be obtained similarly as
described above. As there is no occultation part for the light curve here, 
with a fixed orbital period of 1440 min, fixed inclination, and fixed 
limb-darkening coefficients as before,
only three parameters, $t_{\rm tra}$, $a_{ \rm os}$, $R_{\rm ps }$, are needed for 
the transit model. 
Using the same MCMC sampling process, 
the best-fit model was then obtained with 
a value of reduced chi-square $\chi^2_{\rm red}=4.10$.
The best-fit model and the observational light curve are 
shown in Fig. 7. The MCMC posterior distributions
of these three transit parameters are presented in Fig. 8.

\section{Concluding Remarks}

    To explore the possible existence of exoplanets around brighter stars, 
we searched for exoplanet transits from photometrical data obtained by the BRITE satellite 
through a machine learning technique. Considering exoplanets with orbital periods 
from one to five days, four CNN models focusing on different periods were constructed. 
They were trained with synthetic data generated by injecting planetary transit signals 
into the BRITE light curves.

   Using these four CNN models, i.e. Model P12, P23, P34, and P45, to examine 
the BRITE light curves of 35 stars, 10 stars exhibited possible transits detected 
by at least one CNN models, but only two of them were detected by more than one CNN model. 
These two have larger probabilities to be true planetary transits. Their transit parameters 
were further estimated through an MCMC sampling, and the results show that they could be 
close-in Jupiter-mass exoplanets. The future high-precision observational data will be very 
useful in clarifying whether there are exoplanets in these 10 candidates and also lead to a much 
better determination on the orbital parameters. 
While this study demonstrated that our CNN models could successfully 
identify possible 
transit candidates, we caution that further investigation 
with additional observational data 
is necessary to lead to final confirmations.

\section*{Acknowledgments}
We are thankful to the referee for very helpful suggestions.
This work is based on data collected by the BRITE Constellation 
satellite mission, designed, built, launched, operated and supported 
by the Austrian Research Promotion Agency (FFG), 
the University of Vienna, the Technical University of Graz, 
the Canadian Space Agency (CSA), the University of Toronto 
Institute for Aerospace Studies (UTIAS), 
the Foundation for Polish Science \& Technology 
(FNiTP MNiSW), and National Science Centre (NCN).
This project is supported in part 
by the Ministry of Science and Technology, Taiwan, under
Li-Chin Yeh's 
Grant MOST 106-2115-M-007-014
and Ing-Guey Jiang's
Grant MOST 106-2112-M-007-006-MY3.

\clearpage 
\begin{figure}[ht]
\includegraphics[width=1\textwidth]{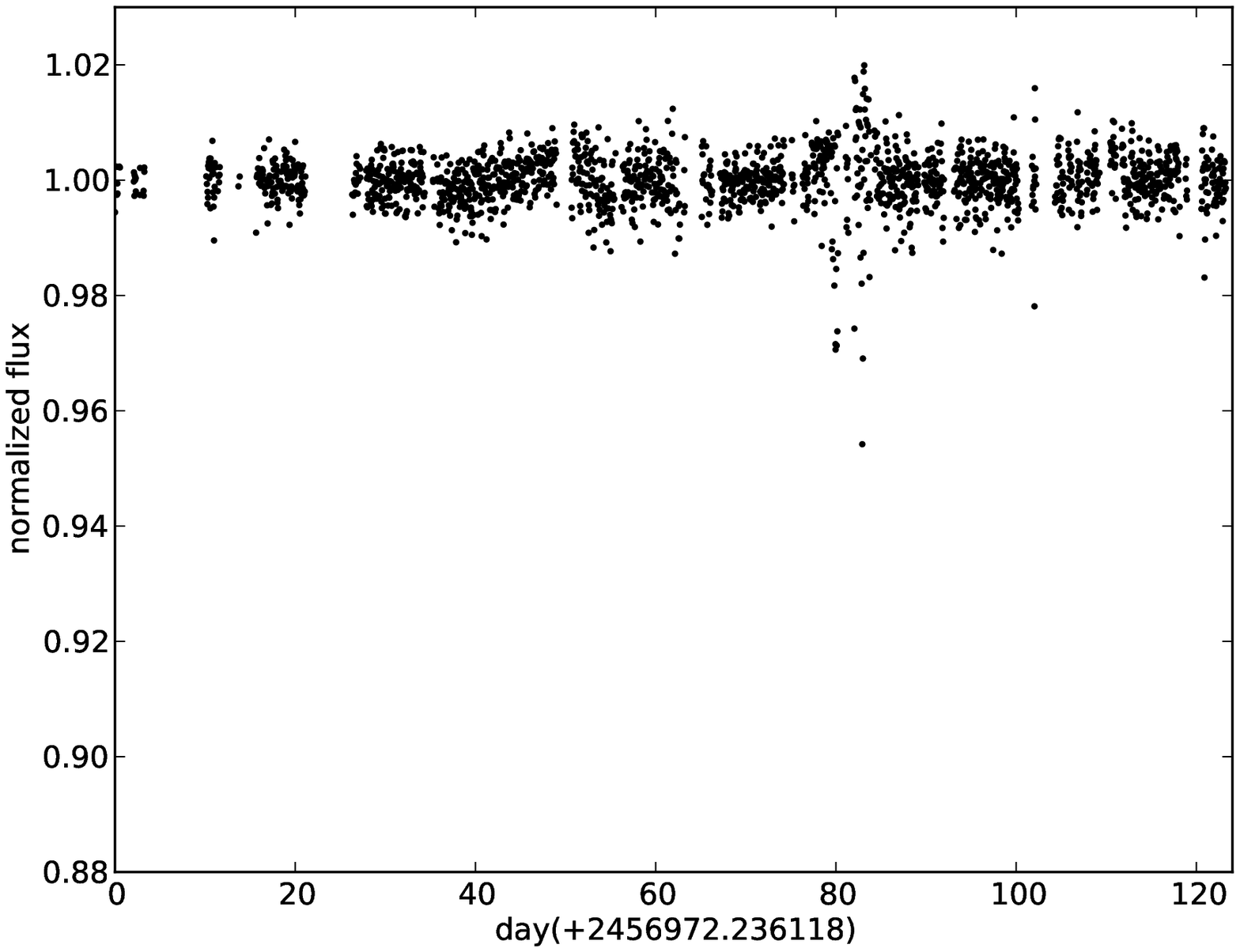}
\caption{The BRITE light curve of HD37468.}
\label{fig1}
\end{figure} 

\clearpage
\begin{figure}[ht]
\includegraphics[width=1\textwidth]{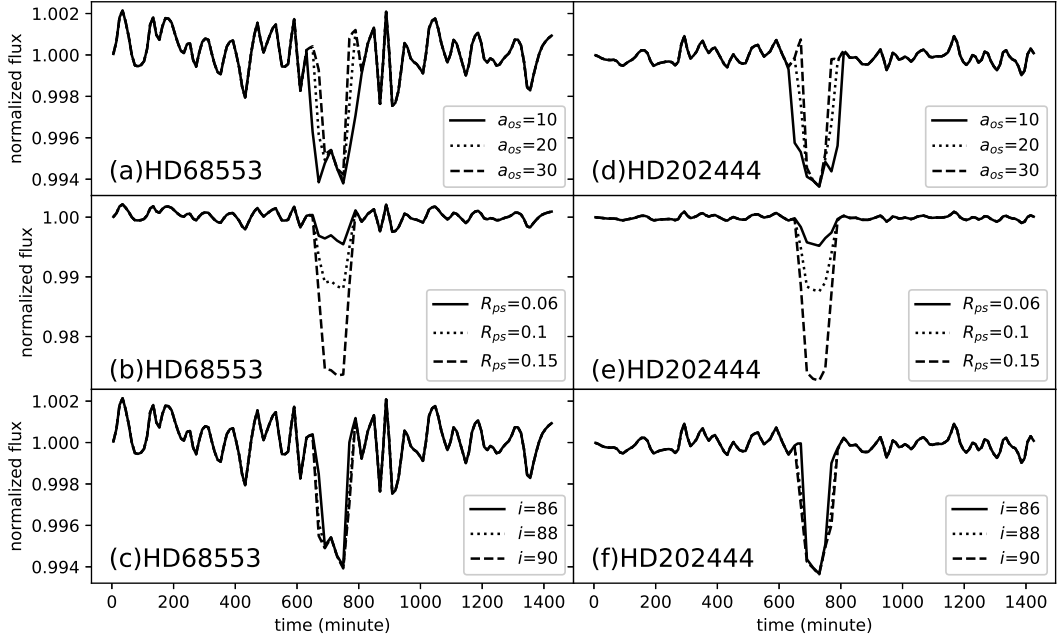}
\caption{The synthetic light curves. 
The left panels represent the star HD68553, and 
the right panels represent the star HD202444.
For top panels, $R_{\rm ps}=0.07$, $i=88$, and 
$a_{\rm os}$=10 (solid curves), 20 (dotted curves), 30 (dashed curves).
For middle panels, $a_{\rm os}=20$, $i=88$, and 
$R_{\rm ps}$=0.06 (solid curves), 0.10 (dotted curves), 0.15 (dashed curves).
For bottom panels, $a_{\rm os}=20$, $R_{\rm ps}=0.07$, and
$i$=86 (solid curves), 88 (dotted curves), 90 (dashed curves). 
} 
\label{fig2}
\end{figure} 

\clearpage
\begin{figure}[ht]
\includegraphics[width=1\textwidth]{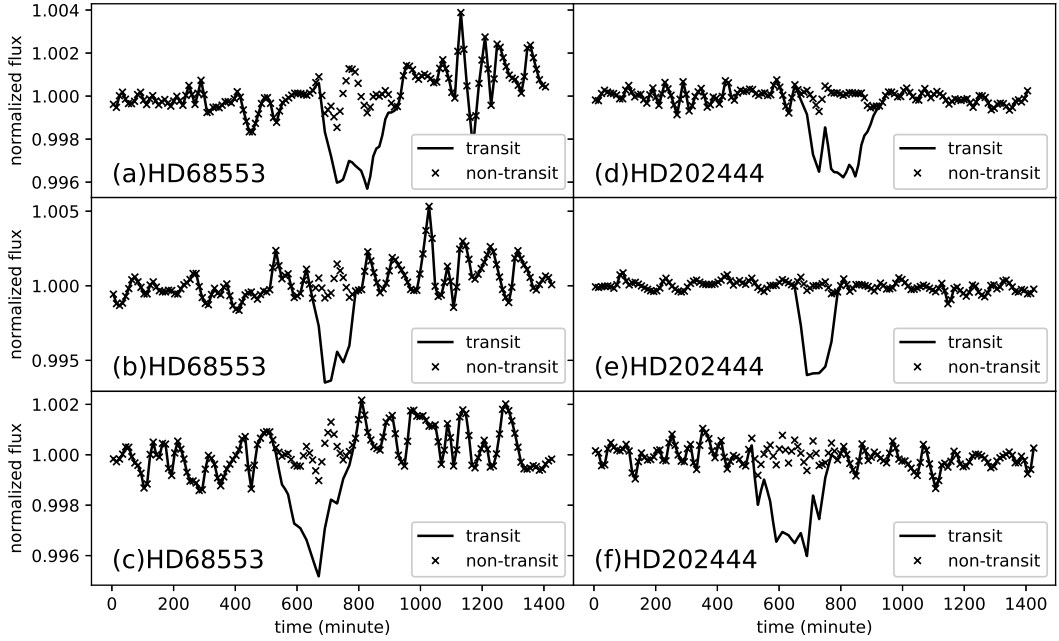}
\caption{The folded light curves, where 
the solid curves are those with transit and the crosses are those 
without transit.
The left panels represent the star HD68553, and 
the right panels represent the star HD202444.
The folding periods for top panels, middle panels, and bottom panels
are 1438 min, 1440 min, and 1442 min, respectively.
} 
\label{fig3}
\end{figure} 

\clearpage
\begin{figure}[ht]
\includegraphics[width=1\textwidth]{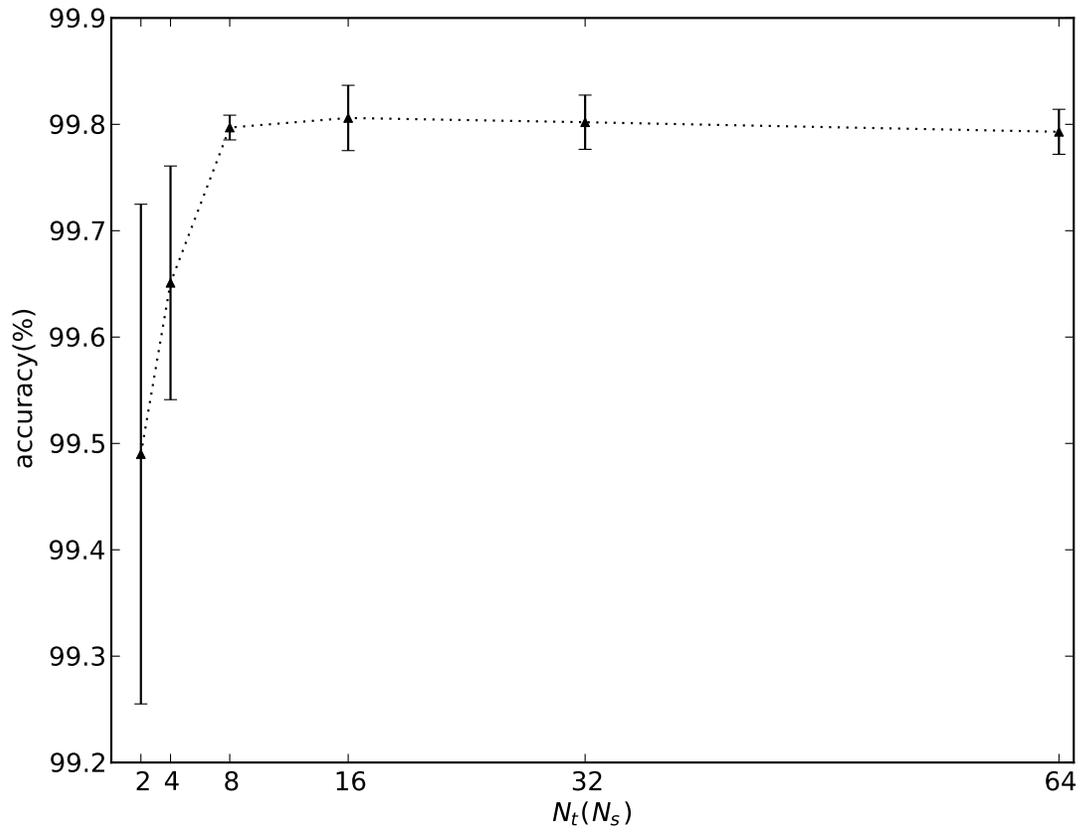}
\caption{The learning curve, representing the accuracy as a function of 
 training sample size.} 
\label{fig4}
\end{figure}

\clearpage
\begin{figure}[ht]
\centering
\includegraphics[width=1\textwidth]{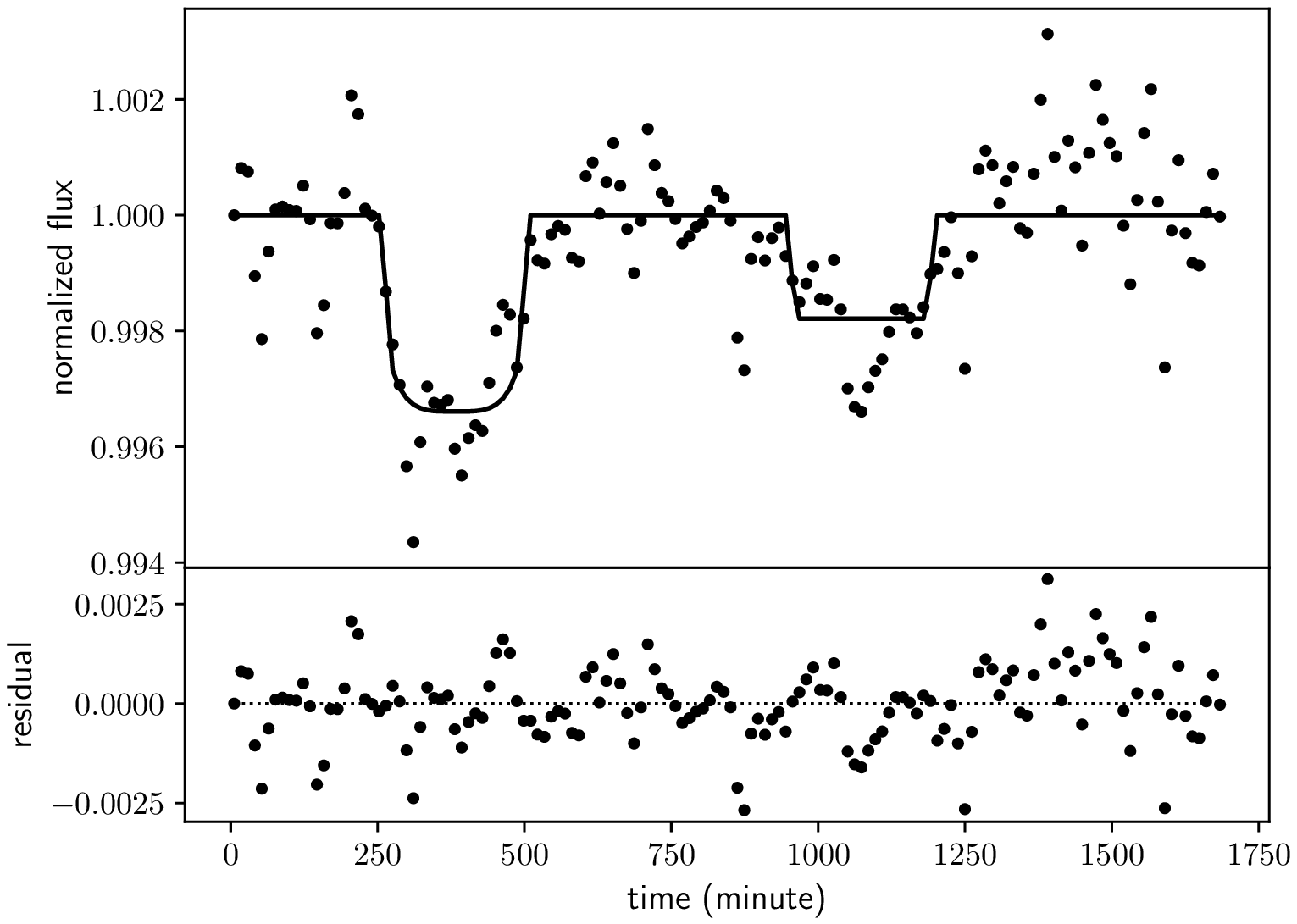}
\caption{The folded BHr-1 light curve of the star HD37468. 
The points are the observational data with the 
 folding period of 1714 minutes. 
The solid line represents the best-fit model.}
\label{fig5}
\end{figure} 

\clearpage
\begin{figure}[ht]
\centering
\includegraphics[width=1\textwidth]{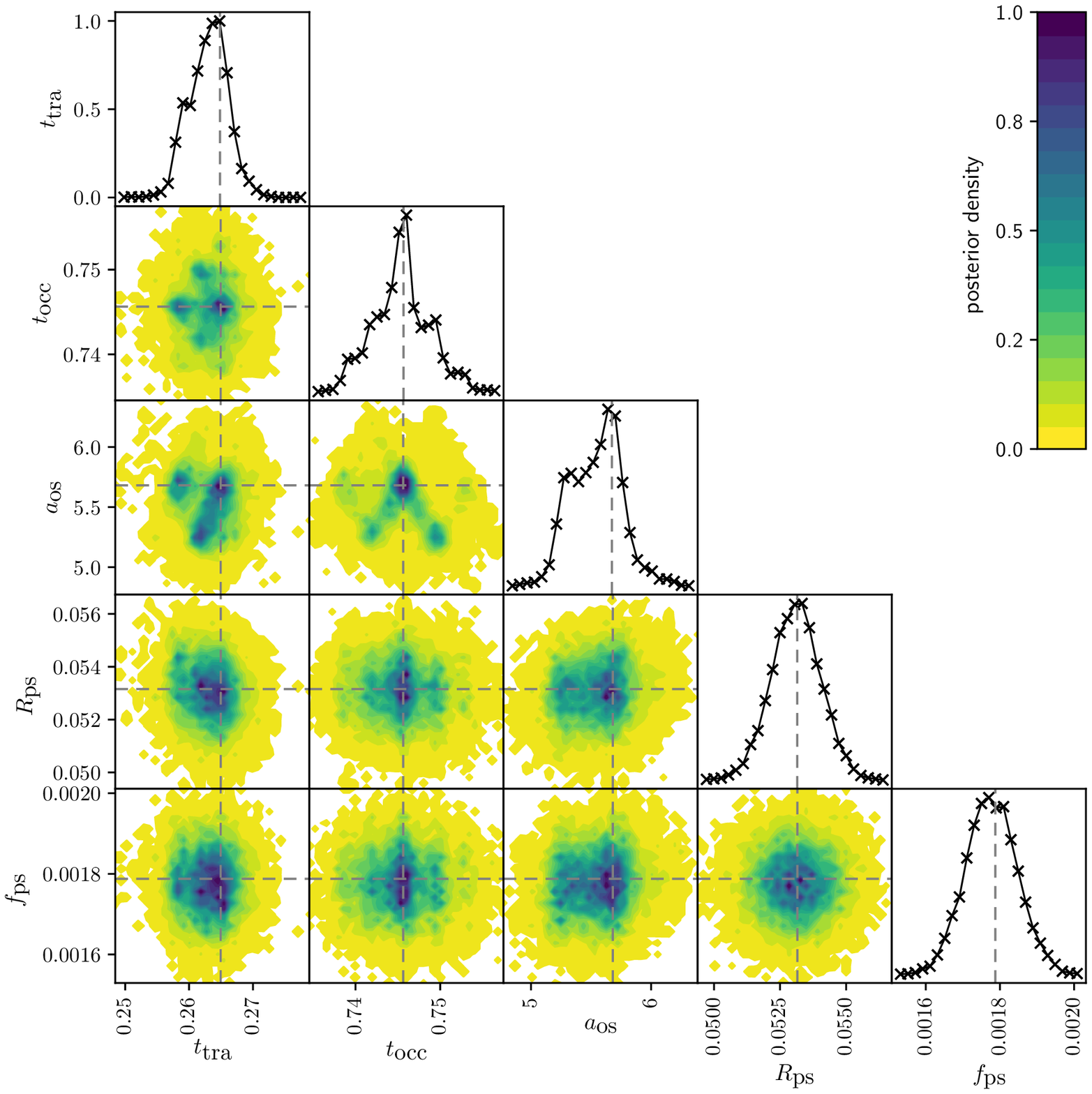}
\caption{The MCMC posterior parameter distributions 
of BHr-1 light-curve modeling of the star HD37468. 
Those panels with colors are pairwise two-dimensional projections. 
The one-dimensional projections
are presented as the histograms on the top.
The dashed lines indicate the parameter 
values of the best-fit model.
}
\label{fig6}
\end{figure}

\clearpage
\begin{figure}[ht]
\includegraphics[width=1\textwidth]{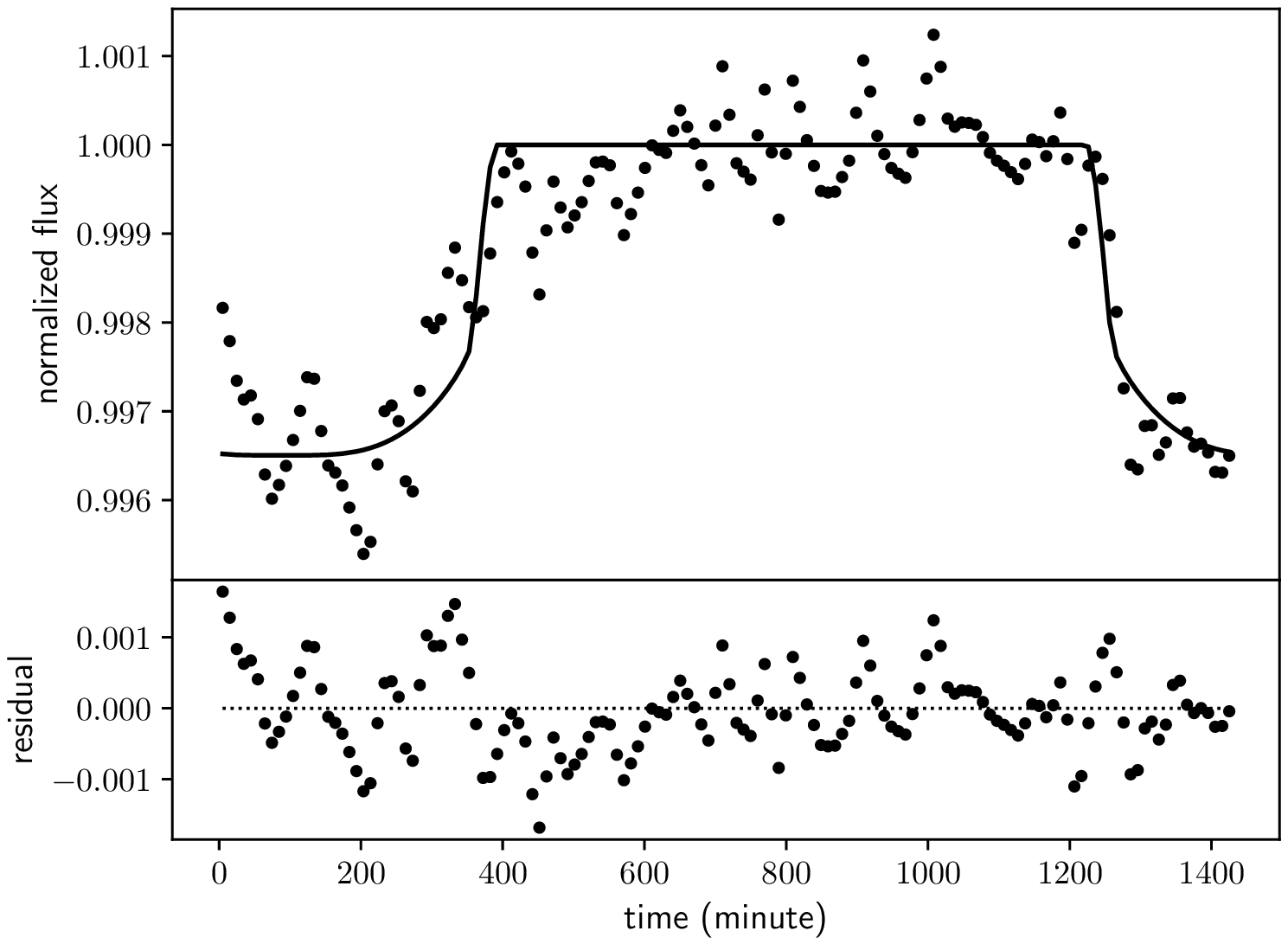}
\caption{The folded BLb-2 light curve of the star HD186882. 
The points are the observational data with the 
folding period of 1440 minutes. 
The solid line represents the best-fit model.
} 
\label{fig7}
\end{figure}  

\clearpage
\begin{figure}[ht]
\centering
\includegraphics[width=1\textwidth]{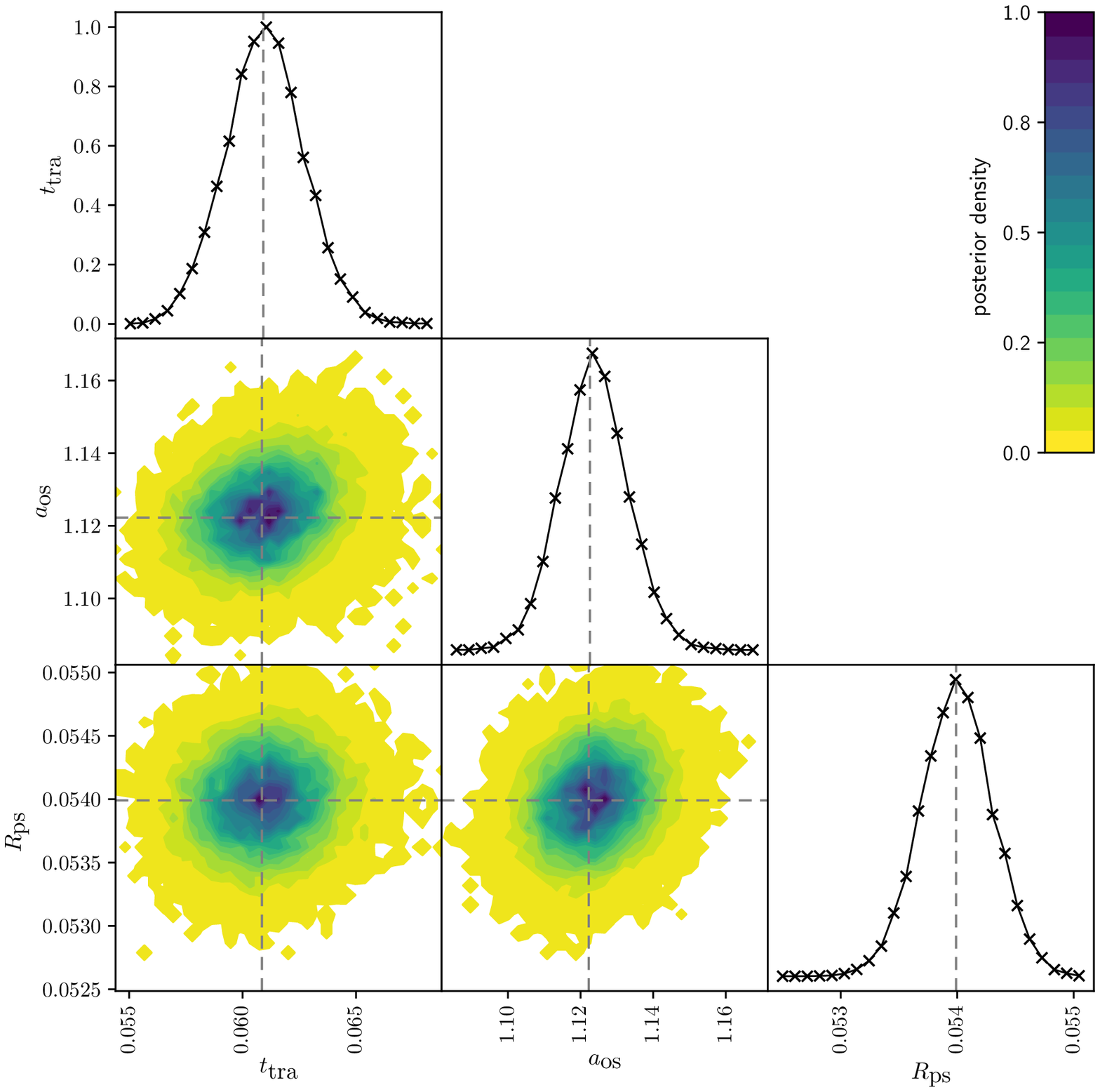}
\caption{The MCMC posterior parameter distributions 
of BLb-2 light-curve modeling of the star HD186882. 
Those panels with colors are pairwise two-dimensional projections. 
The one-dimensional projections
are presented as the histograms on the top.
The dashed lines indicate the parameter 
values of the best-fit model.
}
\label{fig8}
\end{figure}

\end{document}